\documentclass[aps, prb, reprint, twocolumn, superscriptaddress]{revtex4-1}
\usepackage{amsmath}
\usepackage{graphicx}
\usepackage{ dsfont }
\usepackage{dcolumn}
\usepackage{bm}
\usepackage{color,soul}
\usepackage{xcolor}
\usepackage{url}
\usepackage{tabularx}
\usepackage{array}
\usepackage{booktabs}
\usepackage{comment}
\usepackage{hyperref}
\usepackage{footnote}
\usepackage{lipsum}
\usepackage[normalem]{ulem}
\definecolor{colorA}{rgb}{0.006, 0.5, 0.006}

\newcommand{\SUP}[1]{{\color{blue} #1}}

\usepackage{titlesec}
\newcommand*{\justifyheading}{\raggedright}
\titleformat{\section}
  {\normalfont\bfseries\justifyheading}{\thesection}{1em}{}  
\usepackage[singlelinecheck=false, justification=RaggedRight, format=plain]{caption}
\DeclareCaptionLabelSeparator{bar}{ $\boldsymbol{\mid}$ }
\newcommand{\thesupplementbibliography}{\thebibliography}
\usepackage{etoolbox}
\makeatletter
\apptocmd{\thesupplementbibliography}{\global\c@NAT@ctr 30\relax}{}{}
\makeatother

\begin{document}


\title{Multilayer metamaterials with mixed ferromagnetic domain core and antiferromagnetic domain wall structure}

\author{Ruslan Salikhov}
\email{r.salikhov@hzdr.de}
\affiliation{Institute of Ion Beam Physics and Materials Research, Helmholtz-Zentrum Dresden-Rossendorf, Bautzner Landstrasse 400, 01328 Dresden, Germany}

\author{Fabian Samad}
\affiliation{Institute of Ion Beam Physics and Materials Research, Helmholtz-Zentrum Dresden-Rossendorf, Bautzner Landstrasse 400, 01328 Dresden, Germany}
\affiliation{Institute of Physics, Chemnitz University of Technology, Reichenhainer Strasse 70, 09107 Chemnitz, Germany}

\author{Sebastian Schneider}
 \affiliation{Dresden Center for Nanoanalysis, cfaed, Technische Universität Dresden, 01069 Dresden, Germany}
 
\author{Darius Pohl}
 \affiliation{Dresden Center for Nanoanalysis, cfaed, Technische Universität Dresden, 01069 Dresden, Germany}

\author{Bernd Rellinghaus}
\affiliation{Dresden Center for Nanoanalysis, cfaed, Technische Universität Dresden, 01069 Dresden, Germany}

\author{Benny B\"ohm}
\affiliation{Institute of Physics, Chemnitz University of Technology, Reichenhainer Strasse 70, 09107 Chemnitz, Germany}

\author{Rico Ehrler}
\affiliation{Institute of Physics, Chemnitz University of Technology, Reichenhainer Strasse 70, 09107 Chemnitz, Germany}

\author{Jürgen Lindner}
\affiliation{Institute of Ion Beam Physics and Materials Research, Helmholtz-Zentrum Dresden-Rossendorf, Bautzner Landstrasse 400, 01328 Dresden, Germany}

\author{Nikolai~S.~Kiselev}
 \email{n.kiselev@fz-juelich.de}
 \affiliation{Peter Gr\"unberg Institute and Institute for Advanced Simulation, Forschungszentrum J\"ulich and JARA, 52425 J\"ulich, Germany}

\author{Olav~Hellwig}
 \email{o.hellwig@hzdr.de}
\affiliation{Institute of Ion Beam Physics and Materials Research, Helmholtz-Zentrum Dresden-Rossendorf, Bautzner Landstrasse 400, 01328 Dresden, Germany}
\affiliation{Institute of Physics, Chemnitz University of Technology, Reichenhainer Strasse 70, 09107 Chemnitz, Germany}

\date{\today}

\begin{abstract}


Magnetic nano-objects possess great potential for more efficient data processing, storage and neuromorphic type of applications. Using high perpendicular magnetic anisotropy synthetic antiferromagnets in the form of multilayer-based metamaterials we purposely reduce the antiferromagnetic
(AF) interlayer exchange energy below the out-of-plane demagnetization energy, which controls the magnetic domain formation. As we show via macroscopic magnetometry as well as microscopic Lorentz transmission electron microscopy, in this unusual magnetic energy regime, it becomes possible to stabilize nanometer scale stripe and bubble textures consisting of ferromagnetic (FM) out-of-plane domain cores separated by AF in-plane Bloch-type domain walls. This unique coexistence of mixed FM/AF order on the nanometer scale opens so far unexplored perspectives in the architecture of magnetic domain landscapes as well as the design and functionality of individual magnetic textures, such as bubble domains with alternating chirality.

\end{abstract}

\maketitle 

%
The engineering and control of the domain-wall (DW) spin structure in magnetic systems is a crucial area of research for magnetic logic and memory applications, as the dynamic properties of magnetic solitons or DWs themselves strongly depend on it. Additionally, it further impacts the response of DWs to external stimuli, such as magnetic fields and electric currents, and their interaction with other magnetic DWs~\cite{Woo2016,Fert2017,Nagaosa2013,Jiang2017,Litzius2017,Legrand2018,Lucassen2019,Tomasello2014,Malozemoff79,Hubert2009}.
In order to control the internal DW spin structure, it is attractive to use magnetic multilayer (ML) structures, where the respective magnetic energy terms can be adjusted by interfacing
of different materials.
For example, the locking of DW chirality, leading to skyrmion-like magnetic textures, has been realised via the interfacial Dzyaloshinskii-Moriya interaction (IDMI) \cite{Woo2016,Moreau2016,Pollard2017}.
Additionally, by exploiting the Ruderman–Kittel–Kasuya–Yoshida (RKKY) type interaction \cite{Parkin1990,Stiles1999}, domains and DWs that are antiferromagnetically (AF) coupled across a non-magnetic spacer layer have been realized in synthetic antiferromagnets (SAFs)~\cite{Yang2015,Dohi2019,Legrand2020,Chen2020,Juge2022,Chen2022,Duine2018}.
Both the AF-DWs \cite{Yang2015} and skyrmion bubbles in SAFs \cite{Dohi2019} can be driven more efficiently by spin-orbit torques with much smaller current density than it is typically necessary for efficient DW movement in ferromagnetic (FM) MLs. This highlights the potential of SAFs as a promising platform for energy-efficient magnetic logic and memory devices~\cite{Fattouhi2021,Song2020}.
%

Unlike the direct exchange in intrinsic antiferromagnets, the interlayer exchange coupling in SAFs ($E_\text{IEC}$) is much weaker and can be balanced with other magnetic energy terms, such as perpendicular magnetic anisotropy (PMA) ($E_\text{PMA}$) and magnetostatic  dipole interaction ($E_\text{demag}$) \cite{Hellwig2003,Hellwig2004,Hellwig2007,Bran2009,Hauet2008,Hellwig2011}. These terms shape the domain structure and DW spin configuration in magnetic MLs.
The strength of the demagnetization energy in a uniformly magnetized ML system is highly dependent on the orientation of the magnetization with respect to the thin film geometry.
When tuning the interlayer AF exchange coupling energy to be about equal to the out-of-plane demagnetization energy in a PMA-SAF system, both FM and AF phases can coexist~\cite{Hellwig2003,Hellwig2004,Hellwig2007,Bran2009}.
%
%
%
Furthermore, even if the demagnetization energy leads to the formation of a regular periodic FM stripe domain pattern, locally hidden AF configurations will still exist for example within the Bloch type stripe domain walls.
Thus, for a demagnetization energy dominated periodic stripe domain structure with weak AF interlayer exchange coupling, out-of-plane FM domain cores will be separated by in-plane AF Bloch-type domain walls.
This so far unknown mixed FM domain core and AF domain wall structure originates from the unusual maximum energy configuration ($E_\text{PMA}>E_\text{demag}>E_\text{IEC}$), which can be precisely controlled in the multilayer metamaterials that we employ here for its experimental implementation.

%
%

\begin{figure*}[ht]
\centering
\includegraphics[width=17 cm]{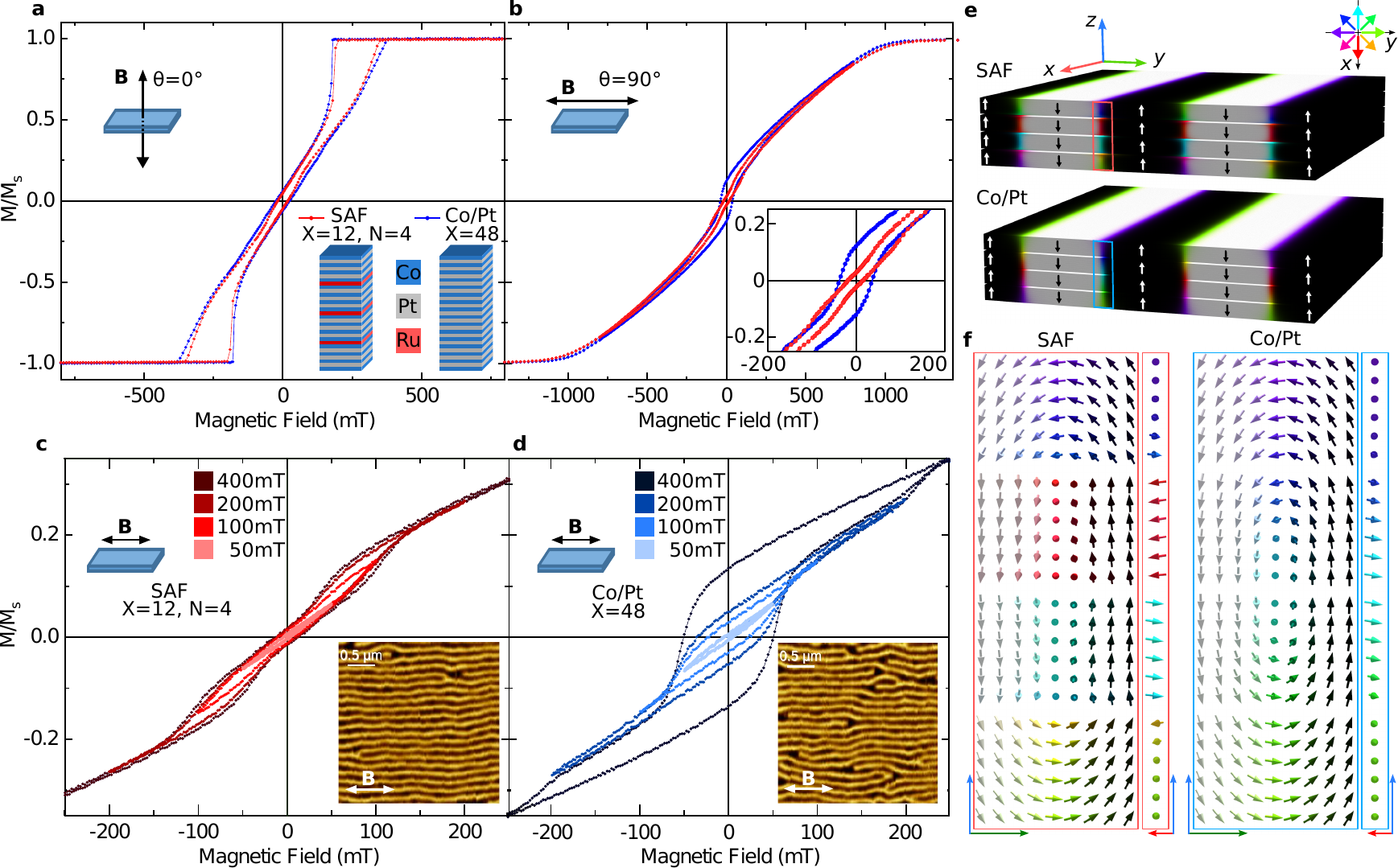}
\caption{\small {\bf Comparison of magnetic states in PMA-SAFs and PMA-FM systems}. \textbf{a}-\textbf{b} Comparison of magnetic hysteresis loops recorded in out-of-plane (\textbf{a}) and in-plane (\textbf{b}) geometry for [[Co(0.44 nm)/Pt(0.7 nm)]$_{11}$/Co(0.44 nm)/Ru(0.9 nm)]$_{3}$[Co(0.44 nm)/Pt(0.7 nm)]$_{12}$ and [Co(0.44 nm)/Pt(0.7 nm)]$_{48}$ MLs. While the out-of-plane loops show identical behaviour, the in-plane loops exhibit different remanent magnetization, as is better seen in an enlarged view in the inset in \textbf{b}. \textbf{c}-\textbf{d} The in-plane minor-loop series with the maximum magnetic field varied between 50 mT and 400 mT for PMA-SAF (\textbf{c}) and PMA-FM (\textbf{d}) systems. The field is applied parallel to the long axis of the stripe domains, as indicated in the  MFM images shown in the corresponding insets.  
\textbf{e} shows the results of micromagnetic simulations for the FM ML (top) and SAF ML (bottom).
Colors encode the direction of magnetization according to a standard scheme: black and white denote up and down spins, respectively, and red-green-blue reflect the direction of the in-plane magnetization component, as shown in the top inset.
\textbf{f} is the zoomed view into the domain walls between up and down domains in SAF ML (left) and FM ML (right). 
\label{Fig1}
}
\end{figure*}

We explore SAF systems using highly tunable, application-friendly ML structures of the type \{[Co/Pt]$_{X-1}$/Co/Ru\}$_{N-1}$[Co/Pt]$_X$, which possess strong PMA. We demonstrate that our PMA-SAF system exhibits FM behaviour with stripe and bubble domains in the ground state, similar to that observed in [Co/Pt]$_{X}$ reference samples.
A detailed study of the DW spin structure and magnetic field behaviour in the PMA-SAFs, however, shows that the Bloch-type DWs possess an AF ordering in-depth. The observation of AF-DWs  between FM domains is consistent with the results of micromagnetic simulations (Fig.\ref{Fig1}\textbf{e,f}). These results extend the spin configuration possibilities for magnetic solitons and provide a way to control and design the internal DW spin structure, which is a key ingredient that defines the dynamic limits for race-track-type applications.

\vspace{0.25cm}
\noindent
\textbf{Multilayer metamaterial system}

In order to demonstrate the key differences between PMA-SAFs and magnetic MLs with PMA, we fabricated two different sets of samples as schematically depicted in the inset of Fig.\ref{Fig1}\textbf{a}. The first set (we refer to as SAF) consists of \{[Co(0.44 nm)/Pt(0.7 nm)]$_{X-1}$/Co(0.44 nm)/Ru(0.9 nm)\}$_{N-1}$[Co(0.44 nm)/Pt(0.7 nm)]$_X$ with $(X,N) = (12,4)$ (or $(10,10)$).
The second set (we refer to as Co/Pt) consists of [Co(0.44 nm)/Pt(0.7 nm)]$_{X}$ PMA-MLs with $X$ = 48 (or 100). The thicknesses of the Co and Pt layers (given in parentheses) were chosen to provide strong PMA (see the Methods section).
The thickness of the Ru-layer was adjusted for the strongest AF coupling between the Co-layers, which has an oscillatory dependence on the Ru thickness due to the RKKY-type interaction~\cite{Parkin1990,Stiles1999}.
The structural comparison of PMA-SAF and PMA-FM systems is provided in the \SUP{Extended Data Fig.1\textbf{a,b}}

\vspace{0.25cm}
\noindent
\textbf{Macroscopic magnetization averaged characterization}

The magnetic hysteresis loops of two samples, SAF $(X,N) = (12,4)$ and Co/Pt $X$ = 48 are compared in Fig.\ref{Fig1}\textbf{a,b}. The out-of-plane magnetization measurements (Fig.\ref{Fig1}\textbf{a}) reveal identical easy-axis behaviour for both samples, indicating that the magnetization reversal mechanism via domain nucleation and annihilation is identical in both systems. Thus, the SAF sample exhibits PMA ferromagnetic characteristics as well. However, the in-plane magnetization measurements (Fig.\ref{Fig1}\textbf{b}), when carefully examined reveal a distinct difference in their hard-axis behavior at remanence (zero magnetic field). Specifically, the Co/Pt sample exhibits a non-zero remanent moment, while the SAF sample has almost no remanence as shown in the inset of Fig.\ref{Fig1}\textbf{b}.
The nonzero remanent moment in Co/Pt is caused by the magnetization within Bloch-type DWs, which separate stripe domains formed at the magnetization reversal, as reported in Ref.[~\cite{Salikhov2021}]. Consequently, the vanishing remanence in the SAF system indicates the absence of the net magnetization from the Bloch-type DWs. A similar behaviour is observed when comparing the SAF $(X,N) = (10,10)$ system with the respective Co/Pt $X$ = 100 ML sample (\SUP{Extended Data Fig.2\textbf{a,b}}).

\begin{figure*}[ht]
\centering
\includegraphics[width=16 cm]{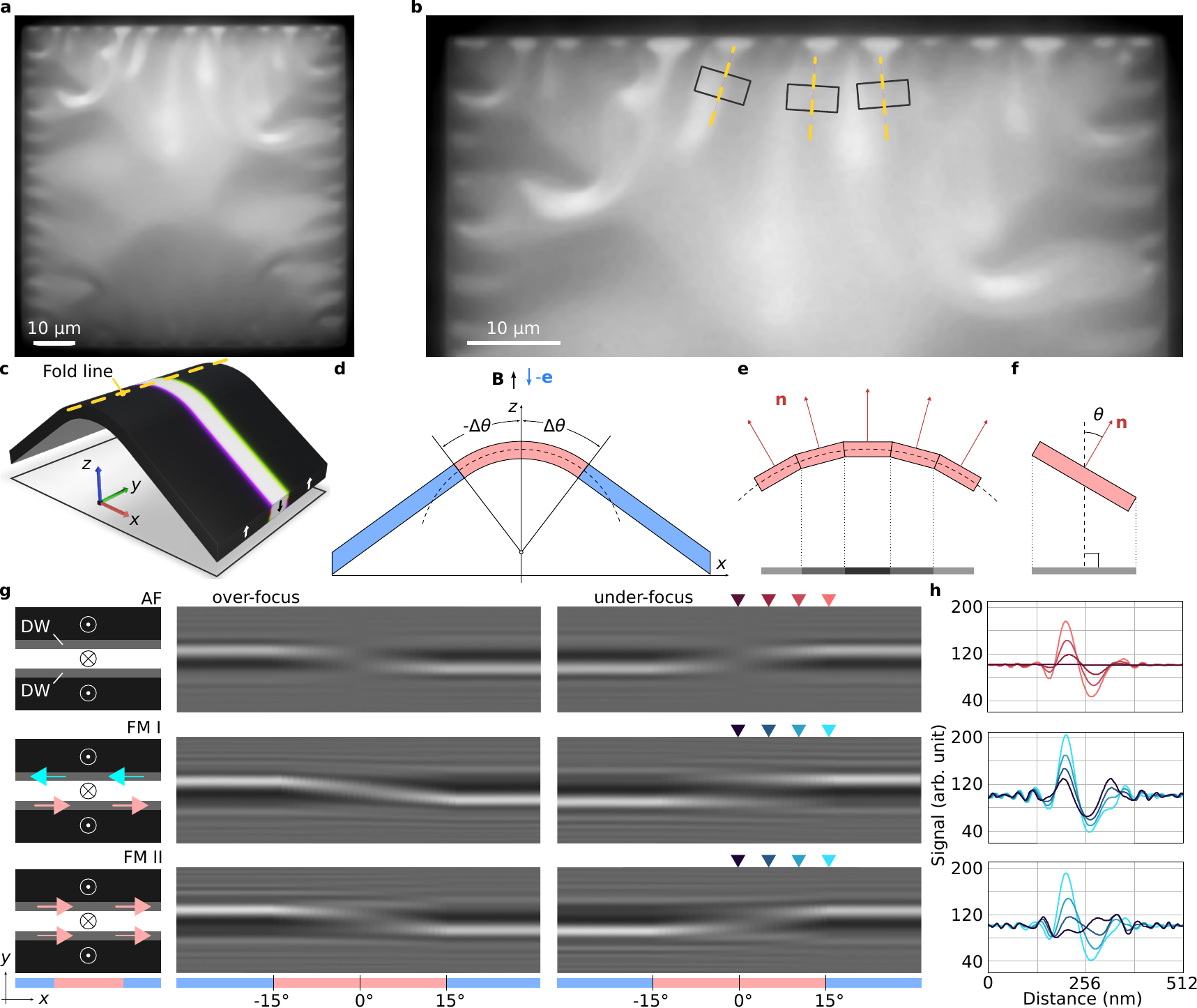}
\caption{\small {\bf Theoretical Lorentz TEM images of isolated stripe domains in PMA-SAF and PMA-FM in the presence of surface folds}. 
\textbf{a} 
Representative example of an experimental in-focus Lorentz TEM image of a ML sample on the Si$_\mathrm{3}$N$_\mathrm{4}$ membrane acquired at low magnification,
\textbf{b} is a magnified area of \textbf{a}. 
The regions with folds appearing due to the buckling of the membrane are marked by black rectangles. The folding lines are marked by (yellow) dashed lines.
\textbf{c} Schematic illustration of a folding line (buckle) in a ML membrane film approximated by a circular arc segment and two flat segments tangential to it. 
In \textbf{d}, the circular and flat segments are marked with red and blue, respectively.
\textbf{e} shows that the arc segment can be effectively approximated by an array of straight segments.
As shown in \textbf{f}, the normal of each straight segment, $\mathbf{n}$, has a certain well-defined tilt angle $\theta$ to the directions of the incident electron beam (blue arrow in \textbf{d}) and external magnetic field (black arrow in \textbf{d}).
\textbf{g} shows theoretical over-focus and under-focus Lorentz TEM images of an isolated stripe domain passing perpendicular through the fold line as illustrated in \textbf{c}.
The top row corresponds to the stripe domains in an SAF, and the two rows below to the stripe domain in a FM Co/Pt ML, with two possible domain wall polarisations, as schematically illustrated in the first column.
The complete images were obtained by placing together the images calculated for flat segments (512 nm $\times$ 512 nm) with tilt angle $\theta$ varying between $-15^\circ$ and $+15^\circ$ in steps of $1^\circ$.
The images were calculated for a defocus of $2.8$ mm.
The plots in \textbf{h} show the cross-section profiles of under-focus images in \textbf{g}. The positions of the cross-sections are marked in \textbf{g} by triangles of the corresponding color. 
\label{Fig2}
}
\end{figure*}

Owing to the fact that the remanent in-plane magnetization originates from the Bloch-type DW polarization, one can study the characteristics of the DW magnetic switching using magnetometry~\cite{Salikhov2021}. To stabilize the aligned stripe-domain state in all samples, an in-plane ac-demagnetization (ACD) protocol is employed (details can be found in the Methods section). After ACD, both systems exhibit parallel stripe domains aligned along the direction of the ACD magnetic field, as evidenced by the magnetic force microscopy (MFM) images in the corresponding insets in Fig.\ref{Fig1}\textbf{c} and \textbf{d}. The domain period of both samples is identical and estimated to be {160 ± 10\,nm}.
Next, we performed a series of minor-loop measurements for SAF (Fig.\ref{Fig1}\textbf{c}) and Co/Pt (Fig.\ref{Fig1}\textbf{d}) samples symmetrically around remanence with the magnetic field applied parallel to the stripe-domain's long axis in order to track the field-dependent DW-magnetization behavior. 
All minor hysteresis loops for the Co/Pt ML, for fields above 50 mT, are hysteretic (Fig.\ref{Fig1}\textbf{d}) with an opening between the ascending and descending branches and an increase in remanent magnetization when the maximal magnetic field is increased. 
This behavior is attributed to the nucleation and propagation of magnetic singularities -- Bloch points of opposite topological charges, resulting in the DW-magnetization switching~\cite{Salikhov2021}. 
In contrast, the SAF system demonstrates almost linear and reversible behavior in response to the applied magnetic field (Fig.\ref{Fig1}\textbf{c}), with no residual magnetization regardless of the maximum field strength. This is a typical characteristic of AF-coupled magnetization. A slight opening at larger magnetic fields can be attributed to the polarization of N\'{e}el-type DW components (see the SAF spin structure in Fig.\ref{Fig1}\textbf{e} and \textbf{f}) through the nucleation of vertical Bloch lines at the top and bottom Co/Pt blocks in the SAF sample. This opening becomes negligible in a thicker SAF system with $(X,N) = (10,10)$, where the N\'{e}el cap volume constitutes a smaller fraction, as evident in \SUP{Extended Data Fig.2\textbf{c}}.

\begin{figure*}[ht]
\centering
\includegraphics[width=17 cm]{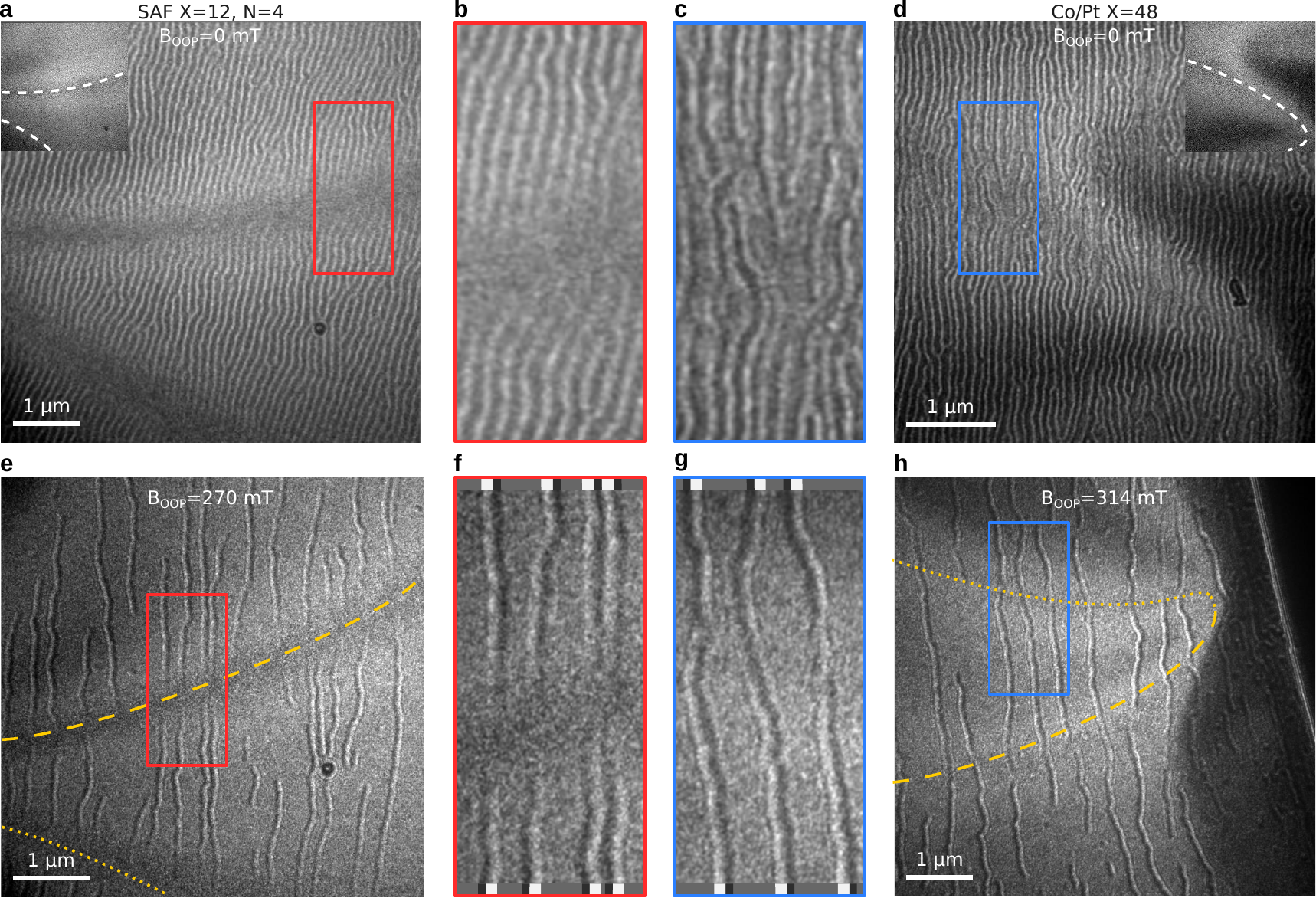}
\caption{\small {\bf Direct evidence for the AF-DW states in SAFs using LTEM in Fresnel configuration}. 
\textbf{a-d} Under-focus Lorentz TEM images of aligned stripe-domain states in SAF (\textbf{a}) and Co/Pt (\textbf{d}) MLs. \textbf{b} and \textbf{c} are magnified areas marked with red and blue rectangles in (\textbf{a}) and (\textbf{d}), respectively.
The images are taken at a defocus of $2$ mm and zero external magnetic field. The insets in (\textbf{a}) and (\textbf{d}) show in-focus Lorentz TEM images of the same area where the fold lines are marked by white dashed lines. 
\textbf{a} illustrates that in the case of SAF, the Lorentz TEM contrast of stripe domains completely vanishes at the fold lines.
In the case of the Co/Pt multilayer in (\textbf{d}), the stripe domains exhibit high contrast, even when crossing the folding line.
The diverse contrast pattern in (\textbf{d}) is because the DWs in Co/Pt sample are polarized in different directions.
In the case of the SAF, in (\textbf{a}), the DWs have zero net magnetization and provide identical contrast patterns.
\textbf{e} and \textbf{h} show the Lorentz TEM images (at a defocus of $2.8$ mm) of isolated stripe domains at a high external magnetic field applied parallel to the e-beam direction in SAF and Co/Pt ML, respectively.
\textbf{f} and \textbf{g} are magnified areas marked with red and blue rectangles in (\textbf{e}) and (\textbf{h}). 
\textbf{f} and \textbf{g}  show an inversion of the contrast of an individual stripe when crossing the fold line.  Note, in \textbf{f}, the contrast completely vanishes at the fold line. Approximate (in-focus) field of view in all images is 7$\mu$m.
\label{Fig3}
}
\end{figure*}

%

Since in our ML system, there are only a few Co layers with asymmetric Pt/Co/Ru interfaces (for the SAF $(X,N) = (12,4)$, the ratio is $6/48$), we exclude any significant contribution of IDMI. Note that IDMI in Pt/Co/Ru and Ru/Co/Pt should have opposite signs, which diminishes the contribution of  IDMI even further.
Additionally, we compared magnetic hysteresis loops in samples with different Ru layer thicknesses as shown in the \SUP{Extended Data Fig.3}. When the Ru thickness is 1.3 nm, where the coupling is FM, the in-plane hysteresis loop exhibits similar characteristics to that of the Co/Pt ML with non-zero remanent magnetization.
However, when we increase the Ru thickness to 1.9 nm, corresponding to the second maximum of the AF-coupling energy, the sample shows a vanishing remanent magnetization, similar to the SAF with 0.9 nm Ru. This provides clear evidence that the AF-like DW behaviour is caused by the AF interlayer exchange coupling and is not caused by any other interfacial effects.

To confirm the DW spin structure in the SAF system, we employed micromagnetic simulations using Mumax~\cite{mumax} and Excalibur~\cite{excalibur}. The details of the micromagnetic simulations are provided in the Methods section.
%
The results of the numerical simulations for the SAF and Co/Pt systems are presented in Fig.\ref{Fig1}\textbf{e}, along with a magnification of the DWs between "up" and "down" domains in Fig.\ref{Fig1}\textbf{f}. Both systems exhibit an identical stripe-domain state, which is in line with the MFM data. 
On the other hand, the simulated DW spin structures of the two systems are very different.
At the film surfaces both, SAF and Co/Pt systems display flux closing N\'{e}el-type DW components.
In the center of the DWs, the Co/Pt system has a FM Bloch-type component which contributes an in-plane net magnetization to the DW.
For the SAF system, the middle of the DW exhibits a subunit-wise in-plane AF ordering of magnetization with zero net magnetization.
This explains the zero net magnetization observed in the in-plane major and minor loop magnetometry measurements in Fig.~\ref{Fig1}\textbf{b}, \textbf{c}.


\vspace{0.25cm}
\noindent
\textbf{Microscopic magnetic characterization of domain walls with Lorentz TEM}

To reveal the presence of AF ordering not only by magnetization averaged measurements but also directly and spatially resolved within individual microscopic DWs, we designed a dedicated experiment using Lorentz transmission electron microscopy (LTEM).
The LTEM measurements are typically conducted on respective sister samples that are deposited onto silicon nitride (Si$_\mathrm{3}$N$_\mathrm{4}$) membrane substrates.
Due to internal mechanical strain between the Si$_\mathrm{3}$N$_\mathrm{4}$ and the metallic MLs these membrane samples usually exhibit buckled profiles with extensive folds and bends that form a characteristic stress pattern~\cite{Ohring} as shown in Fig.~\ref{Fig2}\textbf{a}, \textbf{b}.
In many cases, this aspect is harmful to LTEM imaging, but we use it here specifically to our advantage. In particular, the imaging near the folded areas provides the opportunity to obtain unique information on the magnetic induction of the sample imaged at different angles with respect to the surface normal in one and the same image (Figure~\ref{Fig2}\textbf{c}-\textbf{f}).  
Figure~\ref{Fig2}\textbf{g} illustrates the expected LTEM contrast of an isolated stripe domain pathing across the fold line in the case of SAF and FM samples.
When the electron beam is perpendicular to the membrane surface on top of a fold line, no Lorentz TEM contrast from the DWs in SAF with compensated magnetic moments is detectable (see top row in Figure~\ref{Fig2}\textbf{g}). 
That is because Lorentz TEM is sensitive only to the in-plane components of the magnetic induction.
On the other hand, the projected in-plane magnetization of the DWs in the Co/Pt MLs should result in a pronounced contrast that is consistently present for all angles of the electron beam to the sample normal, also at the top of a fold line, where the beam enters parallel to the surface normal (see the second and third rows in Fig.~\ref{Fig2}\textbf{g}).
Furthermore, the contrast of DWs in Co/Pt multilayers is expected to depend on the polarization direction of the adjacent DWs, as reported in Ref~[\cite{Salikhov2021}].

When a TEM membrane is folded up and down (Fig.\ref{Fig2}\textbf{a} and \textbf{b}) due to the buckling effect,  and the electron beam is adjusted to be parallel to the $z$-axis (Fig.\ref{Fig2}\textbf{c} and \textbf{d}), one obtains signals from the tilted "up" and "down" domain magnetizations where ever the local angle between the electron beam and domain magnetization deviates from zero.  
Since the tilt of the ML film is opposite for sections of the stripe domains on either side of the folding line (i.e., for opposite signs of the slope), their tilt-induced in-plane components of the magnetization have opposite directions. As a consequence the magnetic contrast of the DWs between the stripe domains becomes visible now and is expected to be inverted across the fold line along which the slope is zero. 
%
%
%
For zero slope there is no contrast from the perpendicular domain cores left, here along the fold lines only the DWs contribute to the Lorentz TEM contrast.
Therefore, the inversion behavior of the magnetic contrast from the stripe domains is expected to be different close to the fold line, depending on the DW spin configuration.
According to our simulations, in the case of an AF spin structure, the inversion occurs through zero contrast at the fold line, since the AF-DWs have zero net magnetization (see the first row in Fig.\ref{Fig2}\textbf{g} and \textbf{h}). However, the magnetic contrast from the FM-DWs in Co/Pt MLs contributes to the Lorentz TEM signal across the folding line, resulting in a smooth transition with non-zero contrast, as seen in the second and third rows in Fig.\ref{Fig2}\textbf{g} and \textbf{h}.

\begin{figure*}[ht]
\centering
\includegraphics[width=17 cm]{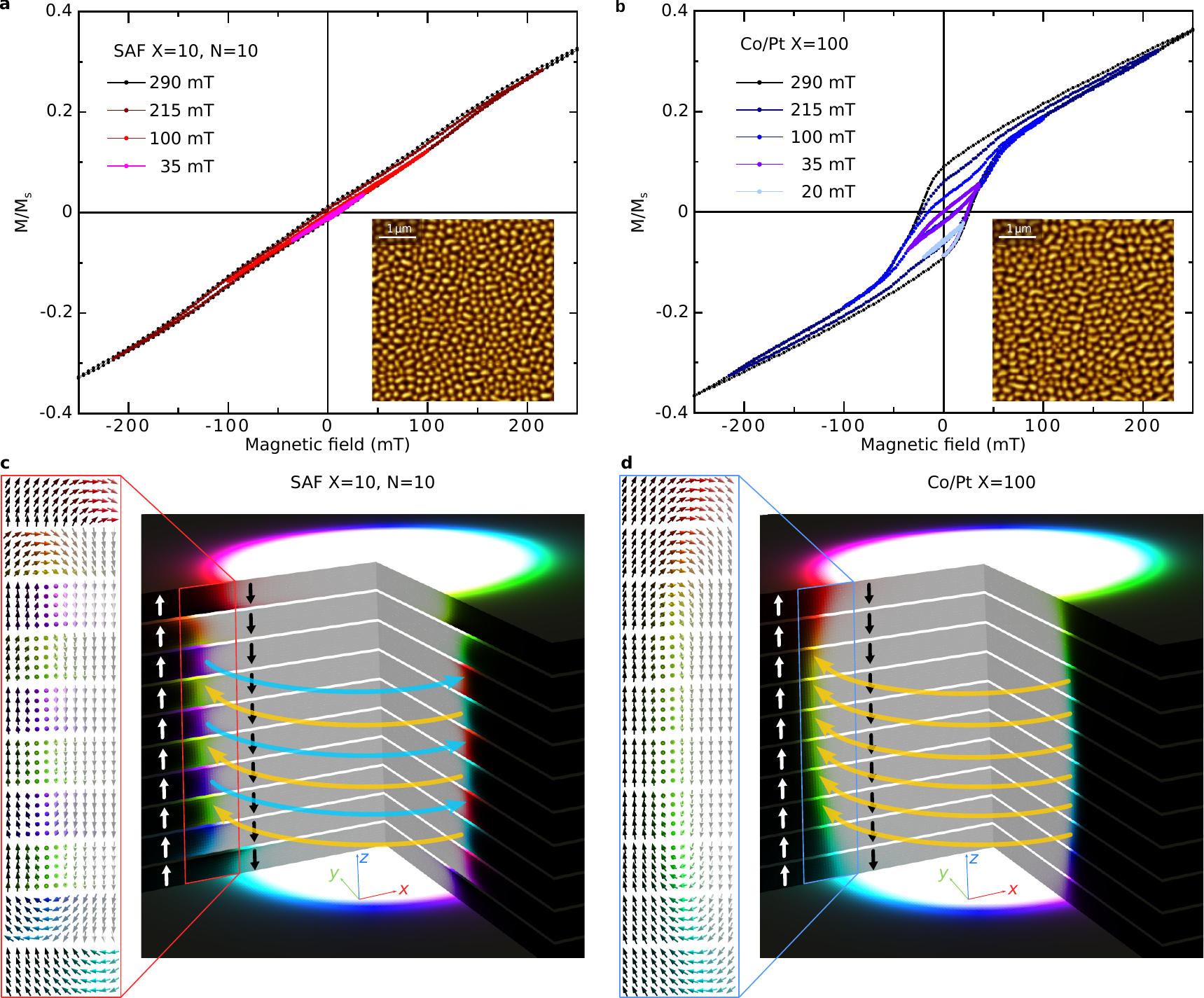}
\caption{\small {\bf Bubble domains in PMA-SAFs and PMA-FM}. 
\textbf{a}-\textbf{b} Magnetic minor hysteresis loop series with the maximum magnetic field varied between 20 mT and 290 mT, for a PMA-SAF [[Co(0.44 nm)/Pt(0.7 nm)]$_{9}$/Co(0.44 nm)/Ru(0.9 nm)]$_{9}$[Co(0.44 nm)/Pt(0.7 nm)]$_{10}$ ML (\textbf{a}) and a PMA-FM [Co(0.44 nm)/Pt(0.7 nm)]$_{100}$ ML (\textbf{b}) system in bubble-domain states. The field is applied parallel to the film plane. Insets represent corresponding bubble-domain states recorded using MFM. \textbf{c}-\textbf{d} The result of micromagnetic simulations for the SAF (\textbf{c}) and FM (\textbf{d}) MLs. The white and black arrows represent the "up" and "down" domain polarization. The spin structure in the corresponding domain walls is color-coded as shown in the zoomed views in the left panels of (\textbf{c}) and (\textbf{d}).    
\label{Fig4}
}
\end{figure*}


By selecting different folding line areas for LTEM imaging (some examples are marked with black rectangles in Fig.\ref{Fig2}\textbf{b}), we were able to obtain representative experimental images of aligned stripe domain states in the SAF and Co/Pt samples, as seen in Fig.\ref{Fig3}\textbf{a} and Fig.\ref{Fig3}\textbf{d} respectively. Figures \ref{Fig3}\textbf{b} and \ref{Fig3}\textbf{c} show a magnification of selected areas in  Fig.\ref{Fig3}\textbf{a,d}. 
Even at large fields of view in Figs.\ref{Fig3}\textbf{a,d} it become apparent that there is a significant difference in the LTEM contrasts between the two FM-domain states. The SAF sample shows no contrast at the fold line (which can be identified from the in-focus TEM image in the inset of Fig.\ref{Fig3}\textbf{a}), while the Co/Pt sample provides magnetic contrast everywhere in the studied area, even at normal electron incidence. This is due to the overlap of signals from the stripe domains in the tilted region with the signal from the Bloch-type DWs. 

To isolate the magnetic signal from a single stripe domain, we apply an external magnetic field of about 270 - 320 mT parallel to the electron beam direction. This causes the "up"-domains to grow and the "down"-domains to shrink.
Figures \ref{Fig3}\textbf{e} and \ref{Fig3}\textbf{h} show the corresponding under focus LTEM images for SAF and Co/Pt samples. Now it becomes apparent that all stripe domains invert the magnetic contrast in the two tilted regions across the folding line, see also the magnified area images in Fig.\ref{Fig3}\textbf{f} and \textbf{g}.
Unlike the Co/Pt sample, which shows a continuous transition due to an additional contribution from Bloch-type DWs, the SAF sample displays no magnetic contrast along the folding line (where the electron beam transmits the sample at normal incidence). The observed behavior of the PMA-SAF sample in the LTEM images confirms on a local level for individual stripe domains the coexistence of mixed FM domain core and AF DW magnetic spin structures and is consistent with our earlier interpretation based on the magnetization averaged minor loop studies.

\vspace{0.25cm}
\noindent
\textbf{Extension from parallel stripe to dense bubble domain states}

Finally, we extend our study to FM bubble domains in SAFs that exhibit AF ordering within their DWs. By using the approach outlined in our previous study~\cite{Salikhov2022}, we stabilized dense bubble states in the $(X,N) = (10,10)$ SAF sample as well as in Co/Pt sample with $X$ = 100. Corresponding MFM images for the SAF and Co/Pt samples are presented in the respective insets in Fig.\ref{Fig4}\textbf{a} and \textbf{b}.  
Without altering the topography of the bubble state we then measured minor hysteresis loop series, similar to those previously performed on the aligned stripe-domain state (see \SUP{Extended Data Fig.2\textbf{c,d}}). 
The results of these measurements for bubble states can be found in Fig.\ref{Fig4}\textbf{a} for the SAF sample and Fig.\ref{Fig4}\textbf{b} for the Co/Pt reference sample. Again, the Co/Pt sample shows irreversible switching of the DWs from one type-II state to another with opposite polarisation via type-I states with different chiralities~\cite{Slonczewski1976,Patek1993}.
The magnetization of the DWs in the SAF sample shows almost fully reversible behavior, akin to antiferromagnets with a linear dependence on the applied field. This behavior is similar to what we previously observed in the aligned stripe-domain state of the same SAF sample with $(X,N) = (10,10)$ (see \SUP{Extended Data Fig.2\textbf{c}}), indicating that the magnetic bubbles in the PMA-SAF are of type-I, with alternating chirality across the Ru-layers.
The latter finding agrees with the results of micromagnetic simulations for the SAF sample in Fig.\ref{Fig4}\textbf{c}, where we also observe the spontaneous formation of alternating chirality in adjacent AF sub-units within the domains walls of the bubble domains.

As seen in Fig.\ref{Fig4}\textbf{c}, the domain walls at the topmost and bottommost layers are shifted with respect to the domain walls in the internal layers. This phenomenon, known as the exchange shift of domain walls, has been studied in the literature earlier~\cite{Kiselev07, Kiselev08}. The exchange shift causes the magnetization in adjacent layers to be ordered antiferromagnetically with respect to each other. In the case of antiferromagnetic IEC, such ordering provides an extra energy gain in the narrow region between the shifted domain walls.

\vspace{0.25cm}
\noindent
\textbf{Conclusions}

Our study reveals the presence of mixed FM/AF textures in SAF MLs with weak RKKY-type interlayer exchange coupling, which is overcome by the strong demagnetization fields within the perpendicular domain cores, but still prevails within the Bloch-type in-plane magnetized domain walls.
As a result, we obtain so far unexplored types of mixed FM/AF textures featuring FM perpendicular stripe and bubble domains separated by AF Bloch-type DWs.
This domain wall structure offers unique possibilities for a wide range of functionalities in PMA-SAFs. Potential applications include using the DWs of aligned stripe domains for nonreciprocal spin wave channeling~\cite{Banerjee2017,Gruszecki2019,Petti2022,Lei2022} and bubble states with  alternating chirality for race track-type applications~\cite{Fert2017,Nagaosa2013,Fattouhi2021}.
Also, recent research has shown that AF-coupled MLs can be reversibly switched between FM and AF states via voltage gating, which could be potentially extended to control the DW state to be either FM or AF in such mixed FM/AF bubble domain arrays~\cite{Kossak2023}.
Finally, also neuromorphic type of computing applications based on magnetic domain systems and textures~\cite{Song2020,Bourianoff2018,Kurenkov2020,Li2021} could benefit from another degree of freedom within the domain wall structure.
Overall our findings extend the perspective for domain wall structure design and pioneer a so far unexplored  pathway towards combining FM and AF order in nanoscale domain structures.

\vspace{0.25cm}
\noindent
\textbf{Methods}

\noindent  \textbf{Sample fabrication.}
The MLs were fabricated using dc magnetron sputter deposition at 0.4 Pa ($3\times10^{-3}$ mbar) Ar atmosphere in an ultrahigh vacuum ATC 2200 system from AJA International Inc. Si wafers with 100-nm-thick thermally oxidized (SiO$_2$) layer as well as Si$_\mathrm{3}$N$_\mathrm{4}$ membranes  were used as substrates. Prior to the ML deposition, a 1.5 nm Ta layer was deposited for adhesion purposes. A subsequent Pt layer (20 nm on SiO$_2$ and 5 nm on Si$_\mathrm{3}$N$_\mathrm{4}$) serves as a seed in order to obtain a preferred Co/Pt (0001)/(111)-texture, which supports better growth and larger PMA~\cite{Fallarino}. The samples were finally capped by a 2 nm Pt layer to avoid surface oxidation.
Structural characterization of the sample architecture has been performed via x-ray reflectivity using a Cu $K_\mathrm{\alpha}$ radiation source diffractometer (Rigaku SmartLab XG). Exemplary reflectivity profiles are displayed for the $(X,N) = (10,10)$ SAF versus $X = 100$ Co/Pt sample in \SUP{Extended Data Fig.1\textbf{a,b}} and allow a clear distinction between the double period ML SAF versus the single period Co/Pt ML.

Magnetic measurements were performed using a commercial Microsense EZ7 vibrating sample magnetometer (VSM), equipped with an electromagnet, which delivers up to 1.8 T magnetic field.
The saturation magnetization of all samples was measured using the VSM and calculated to be $M_\mathrm{s}$ = 0.8 ± 0.1\,MA/m for both SAF and Co/Pt systems. The PMA constant is determined from the in-plane saturation field of a Co/Pt sample with $X$ = 12. Accordingly, by varying the Co thickness and keeping the Pt thickness constant (0.7 nm) we obtained the largest PMA within our sample set to be of about $0.6$\,MJ/m$^3$ at the Co thickness of 0.44 nm. 
The AF exchange constant in SAFs was derived from the calibration
\{[Co(0.44 nm)/Pt(0.7 nm)]$_{X-1}$/Co(0.44 nm)/Ru(0.9 nm)\}$_{N-1}$[Co(0.44 nm)/Pt(0.7 nm)]$_X$ ML with $(X,N) = (5,2)$, which shows a rectangular shape of the field-dependent magnetization in the out-of-plane geometry. 
The interlayer exchange coupling constant $J_\mathrm{iec} \approx -1$\,mJ/m$^2$ was calculated according to the equation $J_\mathrm{iec} = -M_\mathrm{s}d_\mathrm{Co/Pt}B$, where $d_\mathrm{Co/Pt}$ is the thickness of a Co/Pt $X$ = 5 block, and $B$ is the averaged spin-flop field. Having all parameters ($M_\mathrm{s}$, PMA, and $J_\mathrm{iec}$)   for SAF MLs fixed, we varied the $X$ and $N$ until the FM state in SAFs is reached according to the diagram in Ref.~\cite{Hellwig2007}.

The aligned stripe domain states in all the samples were prepared using a specific ac-demagnetization (ACD) routine, namely by alternating the magnetic field direction parallel to the sample surface with subsequent reduction of the field amplitude from 1.8\,T down to 2\,mT in $2\,\%$ steps.
\vspace{5mm}

\noindent  \textbf{Imaging techniques}
Magnetic domain imaging was performed using a magnetic force Bruker Dimension Icon microscope.
The LTEM studies were performed with a JEOL Jem F-200C - transmission electron microscope, operated at 200 kV acceleration voltage, in the Lorentz (Fresnel) mode.
All images were recorded at room temperature.
\vspace{5mm}

\noindent \textbf{Micromagnetic simulations.}
The total micromagnetic energy of the system includes the following terms: Heisenberg exchange, interlayer exchange coupling, uniaxial anisotropy, Zeeman energy, and the energy of the demagnetizing fields.
In a multilayer system, the exchange coupling between Co layers across the Pt inter-layer is weaker than the direct exchange within the Co layer.
We assume that the plane of the multilayer is in $xy$-plane.
Thereby the Heisenberg exchange interaction energy density  within the Co/Pt stack can be approximated as follows~\cite{Salikhov2021}
\begin{equation}
w_\mathrm{ex}=A
\sum_i\!\left[
\left( \dfrac{\partial m_i}{\partial x}\right)^{\!2}\!+\!
\left( \dfrac{\partial m_i}{\partial y}\right)^{\!2}\!+
k_\mathrm{iec}\left( \dfrac{\partial m_i}{\partial z}\right)^{\!2}
\right],
\label{w_ex}
\end{equation}
where the summation runs over $i=x,y,z$ component of the magnetization unit vector field, $\mathbf{m}=\mathbf{M}/M_\mathrm{s}$. The constant $A$ is the exchange stiffness in the plane of the film, while the unitless constant $k_\mathrm{iec}$ defines the strength of the exchange coupling between Co layers across the Pt layers.
In our simulations, we use $A=8$ pJ/m and $k_\mathrm{iec}=0.2$ estimated for [(Co(0.44 nm)/Pt(0.7 nm)]$_X$ multilayers in Ref.~\cite{Salikhov2021}.

The interlayer exchange coupling of Co/Pt stacks across Ru interlayer is defined by the following surface energy density expression:
\begin{equation}
w_\mathrm{iec}=-J_\mathrm{iec}
\left(\mathbf{m}_1\cdot \mathbf{m}_2\right),
\label{w_iec}
\end{equation}
where $\mathbf{m}_1$ and $\mathbf{m}_2$ are the magnetization at the interface, and $J_\mathrm{iec}$ is the coupling constant given in units of J/m$^2$.
For ferromagnetic interlayer exchange coupling, $J_\mathrm{iec}>0$ and for antferromagnetic coupling $J_\mathrm{iec}<0$. 
In practice, on a discrete lattice of cuboids, the energy term \eqref{w_iec} is implemented as follows,
\begin{equation}
w_\mathrm{iec}=-J_\mathrm{iec}
\dfrac{\mathbf{m}_i\cdot \mathbf{m}_j}{\Delta_\mathrm{z}},
\label{fdw_iec}
\end{equation}
where $\mathbf{m}_i$ and $\mathbf{m}_j$ are the magnetization in the cuboids separated by nonmagnetic spacer as illustrated in \SUP{Extended Data Fig.~\ref{Ext_Fig_Theory}\textbf{c}}, and $\Delta_\mathrm{z}$ is the height of the cuboids along $z$-axis.
To implement this interaction in Mumax, we used the option of custom effective fields.
For details of implementation, see the Mumax script file in Supplementary Data.
In our simulations, we use the experimentally estimated value of antiferromagnetic interlayer exchange coupling, $J_\mathrm{iec}=-1$mJ/m$^2$ corresponding to 0.9~-~nm~-~thick Ru layer.
We also use experimentally measured values of saturation magnetization $M_\text{s}~=~775$~kA/m and uniaxial anisotropy $K_\text{u}~=~0.6$~MJ/m$^3$. 
For the calculation of the isolated stripe domain and equilibrium period of stripe domains at zero applied magnetic field (\SUP{Extended Data Fig.~\ref{Ext_Fig_Theory}\textbf{d}}) in $X=12$, $N=4$ multilayer, we use a mesh density of $256 \times 256\times 51$ cuboids. The thickness of the cuboids along the $z$-axis was set to 1 nm, while the lateral size of the cuboid may vary in the range between 1 nm and 2 nm. 
For the simulation presented in Fig.~\ref{Fig1}\textbf{e}-\textbf{f}, we use a domain of square shape in the $xy$-plane with a size of $L_\mathrm{x}=L_\mathrm{y}=320$ nm (twice the stripe-domain period), while for simulations in Fig.~\ref{Fig2}, we use domain with $L_\mathrm{x}=L_\mathrm{y}=512$ nm.  

\vspace{5mm}

\vspace{5mm}
\noindent \textbf{Simulations of Lorentz TEM images}. 
To calculate the Lorentz TEM images in the case of an electron beam perpendicular to the film plane, we follow a well-established approach based on phase object approximation~\cite{MarcDe_Graef}, in detail described in Ref.~\cite{Zheng_21}.
%
%
When the electron beam has an arbitrary tilt angle to the plane of the sample, the magnetic contribution to the phase shift can be calculated as~\cite{Lichte_08}
\begin{equation}
\varphi(x^\prime,y^\prime) = \frac{2\pi e}{h} \int\limits_{-\infty}^{+\infty}\!dz\ \mathbf{A}_\text{d}\cdot \mathbf{n}_\mathrm{e}~, 
\label{phase_shift}
\end{equation}
where $\mathbf{n}_\mathrm{e}$ is the unit vector along the incident electron beam, $e$ is an elementary (positive) charge and $h$ is Planck's constant,
$\mathbf{A}_\text{d}$ is the magnetic vector potential induced by the sample magnetization, and $x^\prime$ and $y^\prime$ are the coordinates in the plane orthogonal to the beam direction. 
%

\vspace{5mm}

\noindent
\textbf{Acknowledgments}

\noindent
The authors are grateful to Thomas Naumann and Jakob Heinze for experimental and technical support. NSK thanks Filipp Rybakov for the fruitful discussions and for providing access to the Excalibur code.

This work was supported by Deutsche Forschungsgemeinschaft through project 514946929 (Gemischt-ferromagnetisch-antiferromagnetische
Hybridphase von Streifen- und "Bubble"-Domänenstrukturen
für magnonische Kristall und "Race-Track"-Anwendungen) and within the priority program SPP 2137 through project 403503416.





\setcounter{figure}{0}
\captionsetup[figure]{labelfont={bf},name={Extended Data Fig.},labelsep=period}

\begin{figure*}[ht]
\centering
\includegraphics[width=14 cm]{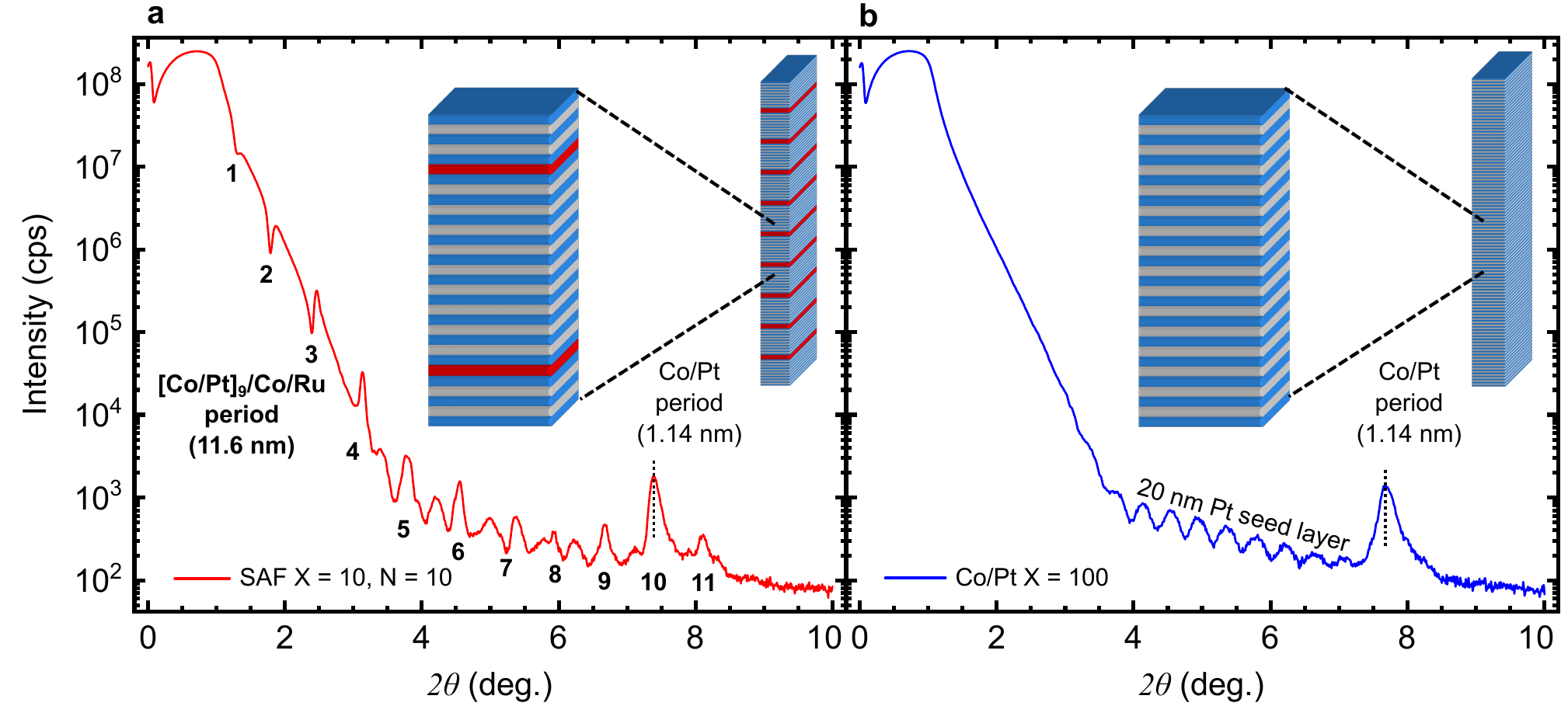}
\caption{\small {\bf Structural comparison of the PMA-SAF and PMA-FM samples}. \textbf{a}-\textbf{b} X-ray reflectivity profiles of the (\textbf{a}) [[Co(0.44 nm)/Pt(0.7 nm)]$_{9}$/Co(0.44 nm)/Ru(0.9 nm)]$_{9}$[Co(0.44 nm)/Pt(0.7 nm)]$_{10}$ PMA-SAF and (\textbf{b}) [Co(0.44 nm)/Pt(0.7 nm)]$_{100}$ PMA-FM samples. 
The top corner insets show the respective schematic layer structure.
The artificial superstructure causes two different types of Bragg reflections for the SAF (\textbf{a}). Up to eleven refraction orders (indicated by bold numbers) are visible for the [Co/Pt]$_{9}$/Co/Ru period (11.6 nm). Higher orders of this structure interfere with the oscillation caused by the 20 nm Pt seed layer, making both hard to distinguish. The Co/Pt bilayer period (1.14 nm) increases the intensity of the 10th order peak (\textbf{b}). The PMA-FM sample shows only the Co/Pt superstructure peak, as marked by the dashed line. The signal from the Pt seed layer at intermediate angles is clearly visible. Both samples display very similar slopes, indicating comparable surface roughness.
\label{Ext_Data_Fig1}
}
\end{figure*}

\begin{figure*}[ht]
\centering
\includegraphics[width=14 cm]{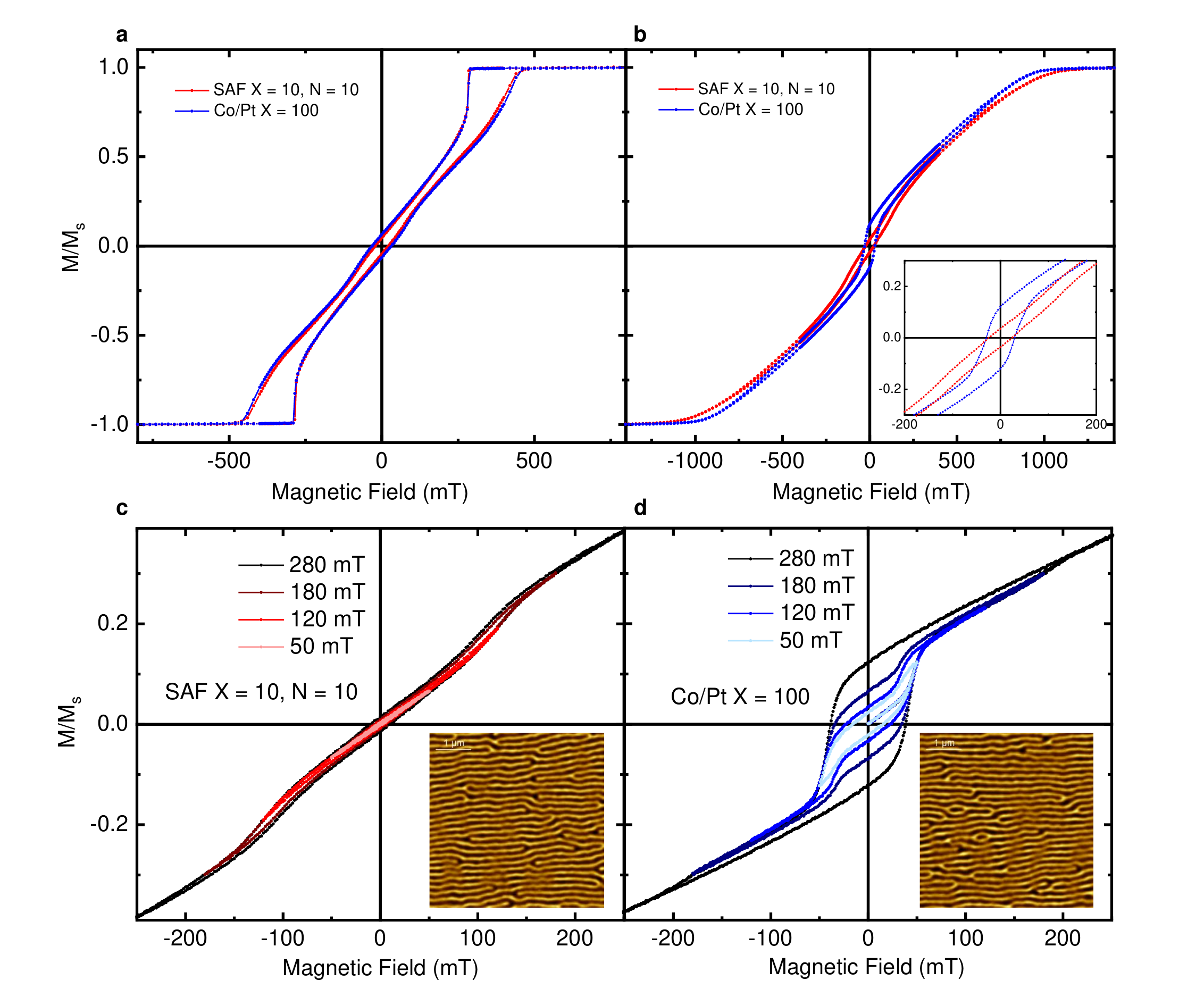}
\caption{\small {\bf Comparison of magnetic states in PMA-SAFs and PMA-FM samples with larger thicknesses}. \textbf{a}-\textbf{b} Comparison of magnetic hysteresis loops recorded in out-of-plane (\textbf{a}) and in-plane (\textbf{b}) geometry for [[Co(0.44 nm)/Pt(0.7 nm)]$_{9}$/Co(0.44 nm)/Ru(0.9 nm)]$_{9}$[Co(0.44 nm)/Pt(0.7 nm)]$_{10}$ and [Co(0.44 nm)/Pt(0.7 nm)]$_{100}$ multilayers. While the out-of-plane loops show identical behavior, the in-plane loops exhibit different remanent magnetization, as is better seen in an enlarged view in the inset in (\textbf{b}). \textbf{c}-\textbf{d} In-plane magnetic minor-loop series with the maximum magnetic field varied between 50 mT and 280 mT for PMA-SAF (\textbf{c}) and PMA-FM (\textbf{d}) systems. The field is applied parallel to the long axis of aligned stripe domains that have previously been stabilized by an ac in-plane demagnetization process. Corresponding MFM images of the parallel stripe domain states are shown as insets in (\textbf{c}) and (\textbf{d}).
\label{Ext_Data_Fig2}
}
\end{figure*}

\begin{figure*}[ht]
\centering
\includegraphics[width=16 cm]{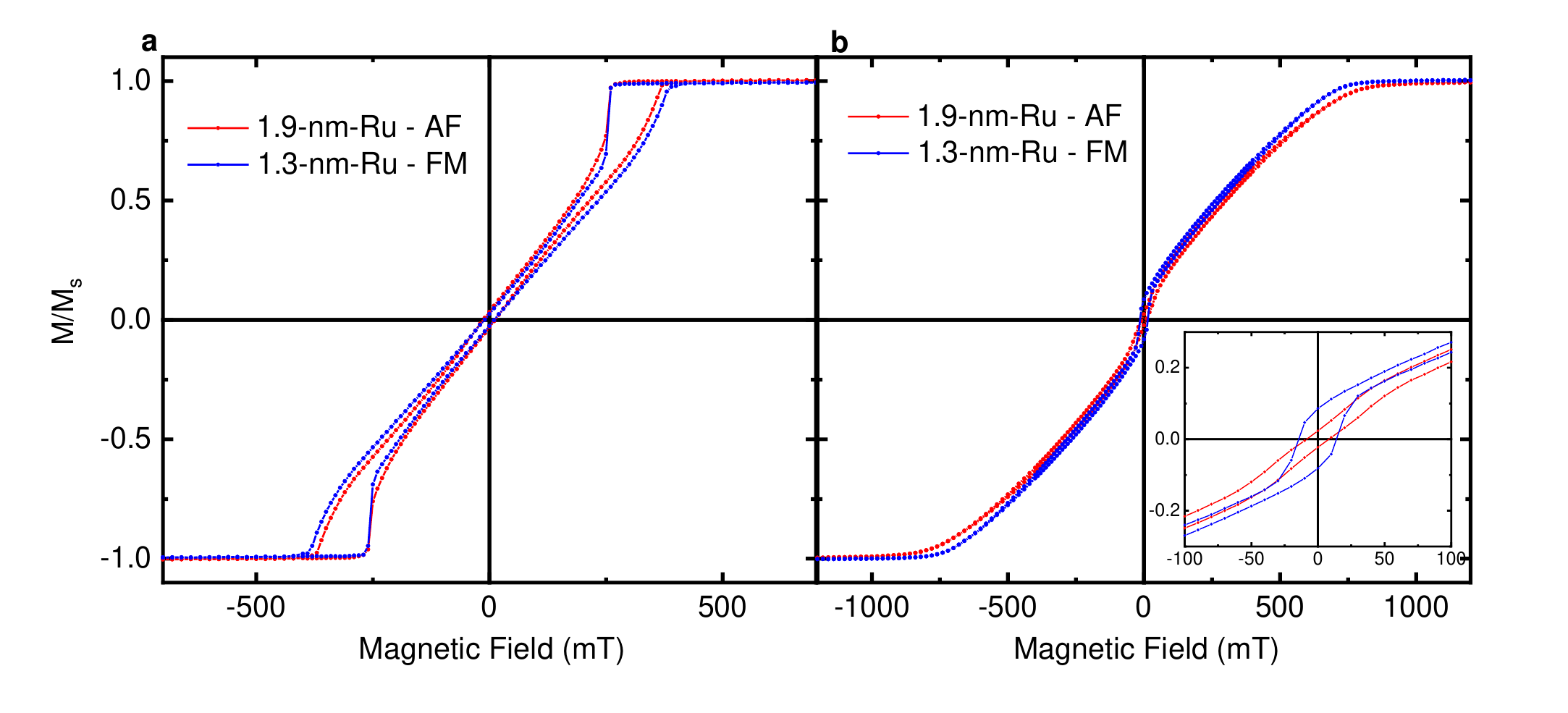}
\caption{\small {\bf Comparison of magnetic states in MLs with different thicknesses of Ru-spacer layer}. \textbf{a}-\textbf{b} Comparison of magnetic hysteresis loops recorded in the out-of-plane (\textbf{a}) and in-plane (\textbf{b}) geometry for [[Co(0.44 nm)/Pt(0.7 nm)]$_{8}$/Co(0.44 nm)/Ru($d$ nm)]$_{17}$[Co(0.44 nm)/Pt(0.7 nm)]$_{9}$ ($(X,N) = (9,18)$) multilayers with different Ru thicknesses $d$ = 1.3 nm and $d$ = 1.9 nm. At the Ru thickness of 1.3 nm the coupling is FM and the in-plane remanent moment is finite (similar to the FM Co/Pt MLs) as is seen in an enlarged view in the inset in (\textbf{b}). At the thickness of 1.9 nm, the second maximum for the AF coupling energy is reached, and the in-plane remanent moment is vanishing (similar to the SAFs with $d$ = 0.9 nm). 
Despite the fact that all interfaces in the PMA-FM and PMA-SAF system are identical, we still observe the same behavior discussed above. This additional experiment confirms that the presence or absence of Co/Ru and Ru/Co interfaces does not significantly impact our findings, which have a more fundamental origin.
\label{Ext_Fig_2}
}
\end{figure*}

\begin{figure*}[ht]
\centering
\includegraphics[width=17 cm]{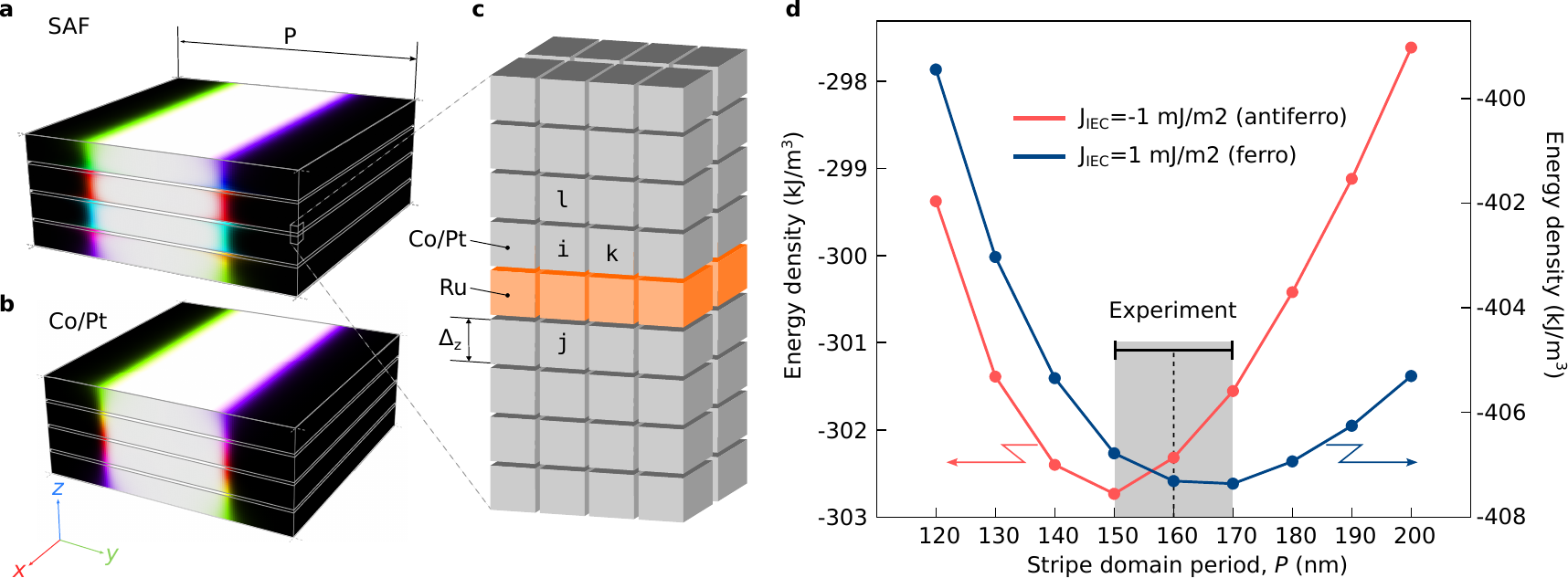}
\caption{\small {\bf Equilibrium stripe domain period for AF and FM interlayer exchange coupling}. \textbf{a}-\textbf{b} show a single period of stripe domain with AF and FM interlayer exchange coupling, respectively.
These states were obtained by direct energy minimization in mumax code assuming periodical boundary conditions in $ xy$-plane.
\textbf{c} illustrates the mesh of cuboids near the Ru interfaces. Cuboids $i$ interact with cuboids $k$ and $l$ within the Co/Pt stack via exchange coupling term \eqref{w_ex} in the Methods section. The interlayer exchange coupling between cuboids $i$ and $j$, sitting next to the nonmagnetic Ru layer, is defined by \eqref{w_iec} and \eqref{fdw_iec}. For details, see the Methods section and Mumax script in Supplementary Data. \textbf{d} Energy density of the system with AF (red curve) and FM (blue curve) interlayer exchange coupling as functions of stripe domains period. The grey rectangle indicates the experimentally estimated periodicity of {160 ± 10\,nm}.
\label{Ext_Fig_Theory}
}
\end{figure*}

\end{document}